\documentclass{llncs}
\usepackage{%
amsmath,%
amssymb,%
mathrsfs,
multirow%
}

\usepackage{algorithm,algorithmic}
\usepackage{a4wide}
\usepackage[pdftex]{graphicx}

\newcommand{\eqdef}{\stackrel{\text{def}}{=}}

\newcommand{\Ft}{\ensuremath{\mathbb{F}_2}}
\newcommand{\Cpub}{{\code{C}_{\text{pub}}}}
\newcommand{\Csec}{{\code{C}_{\text{sec}}}}

\newcommand{\Cr}{\code{C}_{\text{rand}}} 
\newcommand{\Cc}{\code{C}}

\newcommand{\mat}[1]{\ensuremath{\boldsymbol{#1}}}
\newcommand{\code}[1]{\ensuremath{\mathscr{#1}}}

\newcommand{\suppt}[1]{\ensuremath{\textsf{supp}(#1)}}

\newcommand{\prob}{\mathbf{prob}}

\newcommand{\Gm}{\mat{G}}

\newcommand{\Gmp}{\Gm_{\text{pub}}}
\newcommand{\Gms}{\Gm_{\text{sec}}}
\newcommand{\Pm}{\mat{P}}
\newcommand{\Sm}{\mat{S}}

\newcommand{\cv}{\mat{c}}

\newcommand{\ev}{\mat{e}}

\newcommand{\mv}{\mat{m}}

\newcommand{\Ic}{\mathcal{I}}
\newcommand{\Jc}{\mathcal{J}}
\newcommand{\Lc}{\mathcal{L}}

\spnewtheorem{notation}[theorem]{Notation}{\bfseries}{\itshape}
\spnewtheorem{assumption}[theorem]{Assumption}{\bfseries}{\itshape}
\spnewtheorem{fact}[theorem]{Fact}{\bfseries}{\itshape}

\begin{document}
\title{An efficient attack of a McEliece cryptosystem variant based on convolutional codes}
%\author{}
%\institute{}
\author{Gr\'egory Landais and
%\inst{1}
Jean-Pierre Tillich}
\institute{
SECRET Project - INRIA Rocquencourt \\ 
Domaine de Voluceau, B.P. 105   
78153 Le Chesnay Cedex - France \\
\email{gregory.landais@inria.fr}, \email{jean-pierre.tillich@inria.fr}
}

\maketitle

\begin{abstract}
L\"ondahl and Johansson proposed last year a variant of the McEliece cryptosystem which  
replaces Goppa codes by convolutional codes. This modification is supposed to make 
structural attacks more difficult since the public generator matrix of this scheme contains 
large parts which are generated completely at random. They proposed two schemes of this
kind, one of them consists in taking a Goppa code and extending it by adding a generator matrix of 
a time varying convolutional code. We show here that this scheme can be successfully attacked by looking
for low-weight codewords in the public code of this scheme and using it to unravel the convolutional part.
It remains to break the Goppa part of this scheme which can be done in less than a day of computation in
the case at hand.
\medskip

\textbf{Keywords}. Code-based cryptography, McEliece cryptosystem, convolutional codes, cryptanalysis.
\end{abstract}

% -*- mode: latex ; mode: auto-fill; mode: flyspell ; -*-
\section{Introduction}

In \cite{Sho97}, Peter Shor showed that all cryptosystems based on the hardness of factoring or taking a discrete logarithm
 can be attacked in polynomial time with a quantum computer (see \cite{BBD08} for an extensive report). 
This threatens most if not all public-key cryptosystems deployed in practice, such as RSA \cite{RSA78} or DSA \cite{Kra91}. 
Cryptography based on the difficulty of decoding a linear code, on the other hand, is believed to resist quantum attacks and is therefore considered as a 
viable replacement for those schemes in future applications. 
Yet, independently of their so-called ``post-quantum'' nature, code-based cryptosystems offer other benefits even for present-day applications
 due to their excellent algorithmic efficiency, which is up to several orders of complexity better than traditional schemes.

The first code-based cryptosystem is the McEliece cryptosystem \cite{McE78}, originally proposed using Goppa codes.
Afterwards several code families have been suggested to replace the Goppa codes in this scheme:
generalized Reed--Solomon codes (GRS) \cite{Nie86} or subcodes of them \cite{BL05}, Reed--Muller codes \cite{Sid94}, algebraic geometry codes \cite{JM96},
 LDPC codes \cite{BaldiChiara08},  MDPC codes \cite{MTSB12a} or more recently convolutional codes \cite{LJ12a}.
Some of these schemes allow to reduce the public key size compared to the original McEliece cryptosystem while keeping the same level of security against generic decoding 
algorithms. 

However, for several of the aforementioned schemes it has been shown that a description of the underlying 
code suitable for decoding can be obtained- this breaks the corresponding scheme. This has been 
achieved for generalized Reed-Solomon codes in \cite{SidelShesta92} and for subcodes of generalized Reed-Solomon codes
in \cite{Wie10}. In this case, the attack takes polynomial time
and   recovers the complete
structure of the underlying  generalized Reed--Solomon code from the public key $G'$. 
The Reed-Muller code scheme  has also been attacked, but this time
the algorithm recovering the secret description of the permuted Reed-Muller code has sub-exponential complexity \cite{MinderShokrollahi07}
which is enough for attacking the scheme with the parameters proposed in \cite{Sid94} but which is not sufficient to break the scheme
completely. Algebraic geometry codes are broken in polynomial time but only for
low genus hyperelliptic curves~\cite{FM08}. Finally, it should be mentioned that a first version of the scheme based on LDPC codes proposed in \cite{qcldpc3} has been successfully attacked in 
\cite{OTD09} (but the new scheme proposed in \cite{BaldiChiara08} seems to be immune to this kind of attack) and that a variant \cite{BBCRS11a} of the generalized Reed-Solomon scheme which was supposed to resist to the attack of \cite{SidelShesta92}
has recently  been broken in \cite{GOT12a,CGGOT12} by an approach which is related to the distinguisher of Goppa codes
which is proposed in \cite{FGOPT11,FGOPT11a}.

The original McEliece cryptosystem with Goppa codes is still unbroken. It was modified in \cite{BCGO09,MB09} by 
considering quasi-cyclic or quasi-dyadic versions of Goppa codes (or more generally of alternant codes in \cite{BCGO09}) in order
to reduce significantly the key size. However, in this case it was shown that the added structure allows a drastic reduction of the 
number of unknowns in algebraic attacks and most of the schemes proposed in \cite{BCGO09,MB09} were broken by this approach.
This kind of attack has exponential complexity and it can  be thwarted by choosing smaller cyclic or dyadic blocks in this approach in order to increase the number of unknowns of the algebraic system. When the rate of the Goppa code is close to $1$ (as is the case in 
signature schemes for instance \cite{CFS01}) then it has been shown in  \cite{FGOPT11b} that the public key can be distinguished from a random public key.
This invalidates all existing security proofs of the McEliece cryptosystem when the code rate is close to $1$ since they all rely on the hardness of two problems:
 the hardness of decoding a generic linear code on one hand  and the 
 indistinguishability of the Goppa code family on the other hand.

These algebraic attacks motivate the research of
alternatives  to Goppa codes in the McEliece cryptosystem and it
raises the issue of what kind of codes can be chosen in the McEliece cryptosystem.
The proposal with convolutional codes made in \cite{LJ12a} falls into this thread of research.
What makes this new scheme interesting is the fact that its secret generator matrix contains 
large parts which are generated completely at random and has no algebraic structure as in other schemes such 
as generalized Reed-Solomon codes, algebraic geometry codes, Goppa codes or Reed-Muller codes.

In \cite{LJ12a} two schemes are given. The first one simply considers as the secret key the generator matrix of
 a time varying tail-biting convolutional code. A scheme for which it is supposed to resist to attacks of time complexity
 of about $2^{80}$ elementary operations is suggested and has reasonable decoding complexity. This construction presents however the drawback that the complexity
 of decoding scales exponentially with the security level measured in bits. The authors give a second scheme which is scalable and which
 is built upon  a Goppa code and extends it by adding a generator matrix of 
a time varying convolutional code. 

We  study the security of this second scheme in this article. It was advocated that the convolutional structure of the code
can not be recovered due to the fact that the dual code has large enough minimum distance. 
However, we show here that this scheme can be successfully attacked by looking
for low-weight codewords in the public code of this scheme. By a suitable filtering procedure of these low weight codewords 
we can  unravel the convolutional part.

The main point which makes this attack feasible is the following phenomenon : the public code of this scheme contains 
subcodes of much smaller support but whose rate is not much smaller than the rate of the public code. The support
of such codes can be easily found by low weight codewords algorithms. It is worthwhile to notice that the code-based KKS signature scheme 
\cite{kks97} could be broken with exactly the same approach \cite{OT11}. It turns out that the support of these subcodes 
reveals the convolutional structure. By suitably puncturing the public code, there is only the Goppa part which remains. Deciphering an encrypted message can then be done because for the concrete parameters example provided in \cite{LJ12a}, 
algorithms for decoding general linear codes can be used in this case to decode the Goppa code successfully. This attack works successfully
on the parameters proposed in \cite{LJ12a} and needs only a few hours of computation. It should be possible to change the parameters of the
scheme to avoid this kind of attack. In order to do so an improved attack is suggested in Subsection \ref{ss:improvement}, its complexity is analyzed in Section \ref{sec:analysis}. This suggests that it should be possible to repair the scheme by fixing 
the parameters in a more conservative way. Some indications about how to perform such a task are given in Subsection \ref{ss:repair}.

%\input{notation.tex}
% -*- mode: latex ; mode: auto-fill; mode: flyspell ; -*-
\section{The McEliece scheme based on convolutional codes}
\label{sec:scheme}

The scheme can be summarized as follows.

\begin{figure}[!h]
\begin{center}
\includegraphics[scale=0.42]{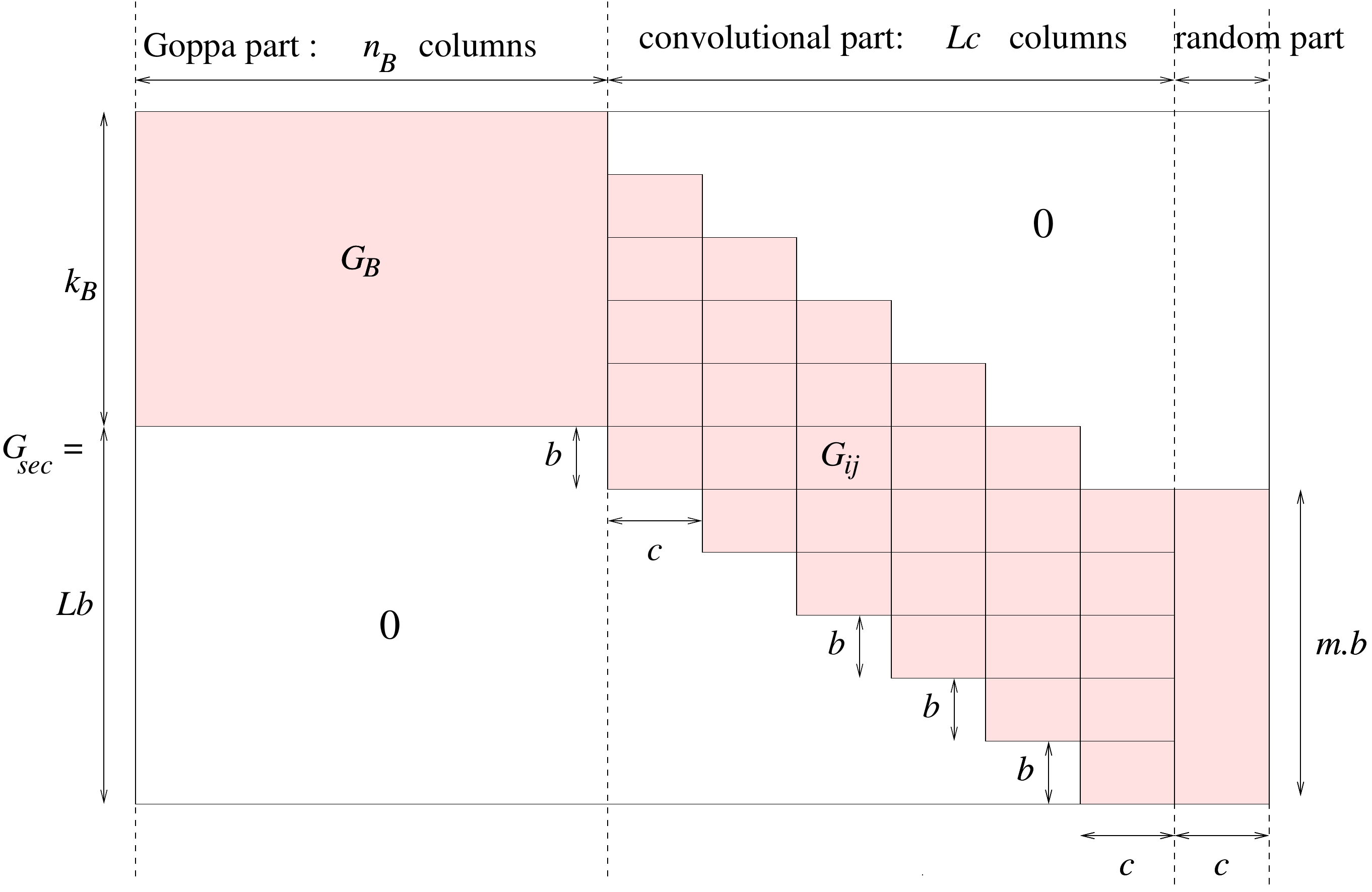}
\end{center}
\caption{The secret generator matrix. The areas in light pink indicate
the only non zero parts of the matrix. $G_B$ is a generator matrix of a
binary Goppa code of length $n_B$ and dimension $k_B$. This matrix is
concatenated with a matrix of a  time varying binary convolutional
code where $b$ bits of information are transformed into $c$ bits of
data (the corresponding $G_{ij}$ blocks are therefore all of size
$b \times c$) and terminated with $c$ random columns at the end. The
dimension
of the corresponding code is $k \eqdef k_B + Lb$ and the length is $n
\eqdef  n_B+(L+1)c$ where $L$ is 
the time duration of the convolutional code.} \label{fig:scheme}
\end{figure}

\begin{description}
	\item \textbf{Secret key.} 
          \begin{itemize}
          \item $\Gms$ is a $k \times n$ generator matrix which has a block form specified in Figure \ref{fig:scheme};
          \item 
            $\Pm$ is an $n \times n$ permutation matrix;
          \item $\Sm$ is a $k \times k$ random invertible  matrix over $\Ft$.
          \end{itemize}
        \item \textbf{Public key.} $\displaystyle \Gmp \eqdef \Sm \Gms \Pm$.
	\item \textbf{Encryption.} The ciphertext $\cv \in \Ft^n$ of a plaintext
          $\mv \in \Ft^k$ is obtained by drawing at random $\ev$
          in $\Ft^n$ of weight equal to some quantity $t$  and computing
          $\displaystyle \cv \eqdef \mv \Gmp  +  \ev$. 
          
	\item \textbf{Decryption.} It consists in performing the following steps
       	\begin{enumerate}
	\item Calculating $\cv' \eqdef \cv \Pm^{-1} = \mv \Sm\Gms + \ev   \Pm^{-1} $
	and using the decoding algorithm of the code with generator matrix $\Gms$  to recover
	$\mv \Sm$ from the knowledge of $\cv'$;
	\item Multiplying the result of the decoding by $\Sm^{-1}$ to recover $\mv$.
	\end{enumerate}
\end{description}

The point of the whole construction is that if $t$ is well chosen,
then with high probability the Goppa code part can be decoded, and 
this allows a sequential decoder of the time varying convolutional
code
to decode the remaining errors. 
From now on we will denote by $\Cpub$ the code with generator matrix
$\Gmp$ and by $\Csec$ the code with generator matrix $\Gms$.

% -*- mode: latex ; mode: auto-fill; mode: flyspell ; -*-
\section{Description of the Attack}
\label{sec:idea}

The purpose of this section is to explain the idea underlying our
attack which is a message recovery attack taking advantage of a
partial
key recovery attack. The attack 
is divided in two  main steps. The first step consists in a (partial) key recovery attack
aiming at unraveling the convolutional structure. The second part consists in a message
recovery attack taking advantage of the fact that if the convolutional part is recovered, then
an attacker can decrypt a message with good probability if he is able
to decode a linear code of dimension $k_B$ and length $n_B$ when there
are less than  $t_{B} \eqdef t \frac{n_B}{n}$ errors (this is the average
number of errors that the Goppa code has to decode).

\subsection{Unraveling the convolutional structure}

\begin{figure}[!h]
\begin{center}
\includegraphics[scale=0.42]{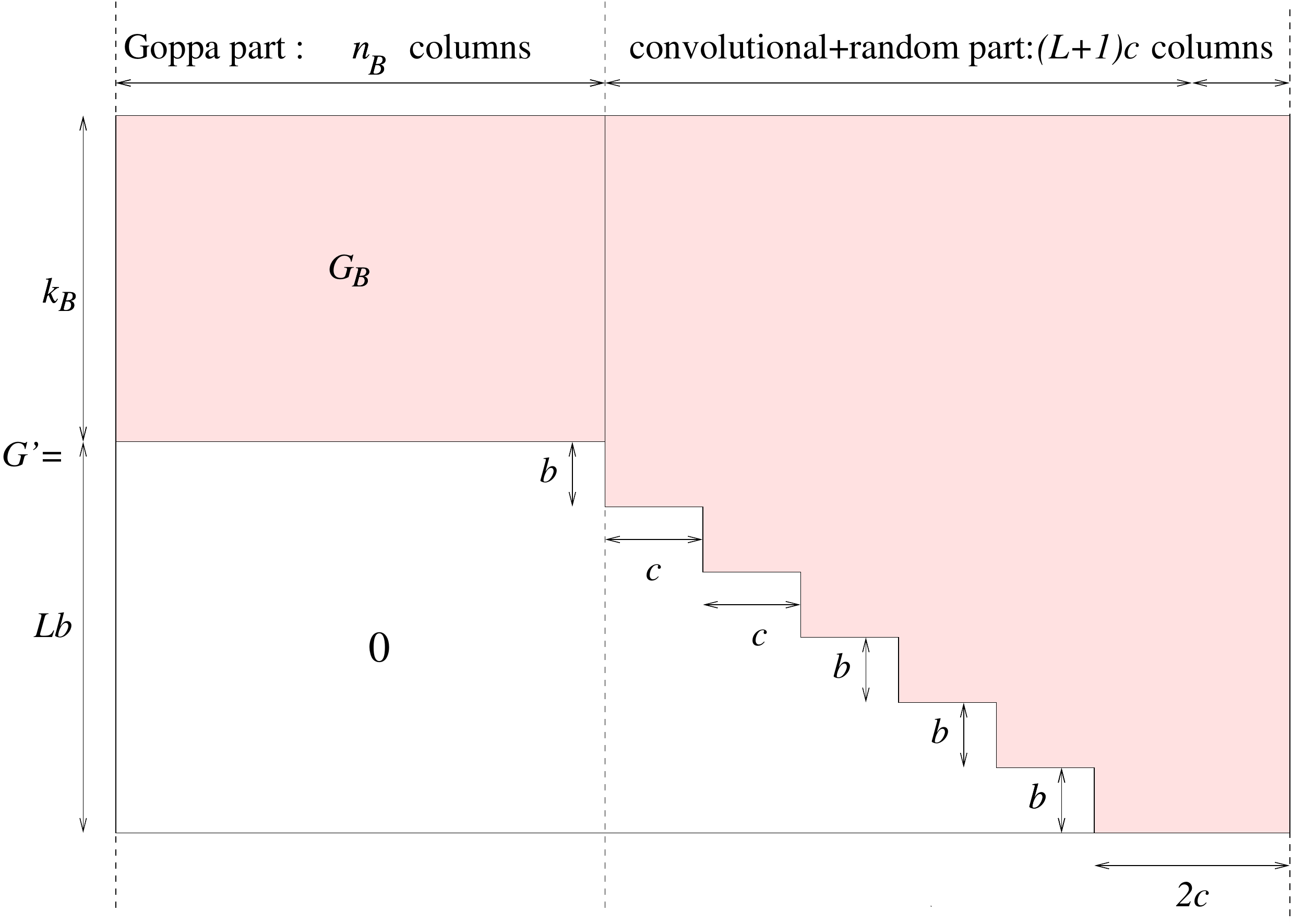}
\end{center}
\caption{The generator matrix of an equivalent code obtained by our
  approach.$\Gm'_B$ denotes the generator matrix of a Goppa code which
is equivalent to the code with generator matrix $\Gm_B$.} \label{fig:equivalent}
\end{figure}

The authors have chosen the parameters of the scheme proposed in
\cite{LJ12a} so that it remains hard to find low-weight codewords in
the dual of the public code $\Cpub$. It is advocated in \cite{LJ12a}
that in their case the only deviation from a random code is the
convolutional structure
in terms of low weight parity-checks.
For instance, the following parameters are suggested
$(n,k)=(1800,1200)$ and in the construction phase the authors propose
to throw away any code who would have parity-checks of weight less
than
$125$. However, the fact that the structure of $\Cpub$ leads in a
natural
way to low weight codewords is not taken into account. Indeed, we
expect many (i.e. about $2^{b-1}$) codewords of weight less than or
equal to $c$. This comes from the fact that the subcode of $\Cpub$
generated by the last $b$ rows of $\Gms$ (and permuted by $\Pm$)
has support of size $2c$ and dimension $b$. Therefore any algorithm
aiming at finding codewords of weight less than $c$ say should output 
such codewords. Looking at the support  of such codewords reveals the $2c$ last
columns
of $\Gms$. By puncturing these columns and starting this process again
but this time by looking for codewords of weight less than $c/2$
(since this time the punctured code 
contains a subcode of dimension $b$ and support of size $c$ arising 
from the penultimate block of rows of $\Gms$) will reveal the
following block of $c$ columns of the matrix. 
In other words we expect to capture by these means a first subcode
of dimension $b$ and support the $2c$ last positions of $\Csec$.
Then we expect a second subcode of dimension $b$ with support the
$3c$ last positions of $\Cpub$ and so on and so forth.
Finally we expect to be able after suitable column swapping to 
obtain the generator matrix $G'$ of an equivalent code to $\Cpub$ which
would have the form indicated in Figure \ref{fig:equivalent}.

More precisely the algorithm for finding a generator matrix of a code
equivalent to $\Cpub$ is given by Algorithm \ref{algo:convolutional} given below.

\begin{algorithm}
  \caption{An algorithm for finding $\Gm'$. \label{algo:convolutional}}
  {\bf input:} $\Gmp$ the public generator matrix\\
%  \Comment{$\xb^1,\dots,\xb^M$ sont les parties d'information bruitées des $M$ mots de code re\c{c}us et
%les $\zb^1,\dots,\zb^M$ sont leurs redondances bruitées (par rapport au second code convolutif).}
{\bf  output:} a generator matrix $\Gm'$ of a code equivalent to $\Cpub$ which has the form indicated in 
Fig. \ref{fig:equivalent}.\\
\begin{algorithmic}
\STATE{$\Lc \leftarrow []$}
  \FOR{$i=L,\ldots{},1$}
  \STATE{$\Gm \leftarrow \text{{\tt GeneratorMatrixPuncturedCode}}(\Cpub,\Lc)$}
  \STATE{$\Gm \leftarrow \text{{\tt LowWeight}}(\Gm,w)$}
  \STATE{$w \leftarrow \text{{\tt Function}}(i)$}
  \STATE{$\Gm_{i}\leftarrow \text{{\tt ExtendedGeneratorMatrix}}(\Gm,\Lc,\Cpub)$}
   \STATE{$ \Lc \leftarrow \text{{\tt Support}}(\Gm)|| \Lc$}
  \ENDFOR
  \STATE{$\Gm \leftarrow \text{{\tt GeneratorMatrixPuncturedCode}}(\Cpub,\Lc)$}
   \STATE{$\Gm_{0}\leftarrow \text{{\tt ExtendedGeneratorMatrix}}(\Gm,\Lc,\Cpub)$}
\STATE{$\Gm'$ is the concatenation of the rows of $\Gm_{0},\Gm_{1},\ldots,\Gm_{L}$.}
  \RETURN $\Gm'$ 
  \end{algorithmic}
\end{algorithm}

We assume here that :
\begin{itemize}
	
	\item the function {\tt GeneratorMatrixPuncturedCode}  takes as input a code $\code{C}$ of length $n$ and an ordered set of positions $\Lc$ which is a sublist of $[1,2,\dots,n]$ and 
outputs a generator matrix of $\code{C}$ punctured in the positions belonging to $\Lc$;
\item  \text{{\tt Function}} will be a certain function which will be specified later on;
\item \text{{\tt Support}}$(\code{C})$ yields the (ordered) support of $\code{C}$ and $||$ is the concatenation of lists;
\item the function {\tt LowWeight} takes as input a code $\code{C}$ and a weight $w$. It
outputs a generator matrix of a subcode of $\code{C}$ obtained by looking for codewords of weight less than or equal to 
$w$.
Basically a certain number of
codewords of weight $\leq w$ are produced and the positions which are involved in at least  $t$ codewords 
are put in a list $\Lc$ (where $t$ is some threshold
depending on the weight $w$, the length $n$ of the code, its dimension $k$ and the number of
codewords produced by the previous algorithm), which means that $i$ is taken as soon as there are at least 
$c$ elements in $\Cc$ for which $c_{i}=1$. Then a generator matrix for
the subcode of $\code{C}$ formed by the codewords of $\code{C}$ whose coordinates outside
$\Lc$ are all equal to $0$ is returned. See Algorithm \ref{algo:lowweight} for further details.

\item the function {\tt ExtendedGeneratorMatrix} takes as input a generator matrix of some code $\code{C'}$, an ordered set of positions 
$\Lc$ and a code $\code{C}$ such that $\code{C'}$ is the result of the puncturing of  $\code{C}$  in the positions belonging to $\Lc$. It
outputs a generator matrix of the permuted subcode $\code{C''}$ of $\code{C}$ whose positions are reordered in such a way that the first positions correspond to the positions of $\code{C'}$ and the remaining positions to the ordered list $\Lc$. This
code $\code{C''}$ corresponds to the codewords of $\code{C'}$ which are extended as codewords of $\code{C}$ over 
the positions belonging to $\Lc$ in an arbitrary linear way.
\end{itemize}

\floatname{algorithm}{Algorithm}
\begin{algorithm}
	\caption{ {\tt LowWeight}$(\Gm,w)$ \label{algo:lowweight}}
  {\bf input:}
  \begin{itemize}
  \item
   $\Gm$ a certain $k \times n$ generator matrix of a code $\code{C}$;
   \item
  $w$ a certain weight.
  \end{itemize}
%  \Comment{$\xb^1,\dots,\xb^M$ sont les parties d'information bruitées des $M$ mots de code re\c{c}us et
%les $\zb^1,\dots,\zb^M$ sont leurs redondances bruitées (par rapport au second code convolutif).}
{\bf  output:} a generator matrix $\Gm'$ of a subcode of $\code{C}$ 
obtained from the supports of a certain subset of codewords of weight $w$ in $\code{C}$. \begin{algorithmic}
%\STATE{produce a set $\Cc$ of linear combinations of rows of $\Gm$ of weight less than $w$ by running a low-weight codeword search algorithm.}
\STATE{$\Cc \leftarrow \text{{\tt LowWeightCodewordSearch}}(\Gm, w)$}
\COMMENT{Produces a set of linear combinations of rows of $\Gm$ of weight~$\leq~w$}
\STATE{Initialize an array $tab$ of length $n$  to zero}
\STATE{$t \leftarrow \text{{\tt Threshold}}(w,n,k,|\Cc|)$}
\FORALL{$c \in \Cc$}
\FOR{$i \in [1..n]$}
\IF{$c_i=1$}
\STATE{$tab[i] \leftarrow tab[i]+1$}
\ENDIF
\ENDFOR
\ENDFOR
\STATE{$\Lc \leftarrow []$}
\FOR{$i \in [1..n]$}
\IF
{$tab[i] \geq t$}
\STATE{$\Lc \leftarrow \Lc || \{i\}$}
\ENDIF
\ENDFOR
\STATE{$\Gm' \leftarrow \text{{\tt SorthenedCode}}(\Gm,\Lc)$}
\COMMENT{Produces a generator matrix for
the subcode of $\code{C}$ formed by the codewords of $\code{C}$ whose coordinates outside
$\Lc$ are all equal to $0$.}
\RETURN $\Gm'$.
\end{algorithmic}
\end{algorithm}

\subsection{Finishing the job : decoding the code with generator
  matrix $\Gm'_B$}

If we are able to decode the code with generator matrix $\Gm'_B$, then
standard sequential decoding algorithms for convolutional codes will
allow to decode the last $(L+1)c$ positions. Let $\Gm_{B}'$ be the
generator matrix of a code equivalent to the secret Goppa code chosen for the scheme
specified in Figure \ref{fig:equivalent}. Decoding such a code can be done by 
algorithms aiming at decoding generic linear codes such as Stern's algorithm \cite{Stern88} and its
subsequent improvements \cite{Dumer91,BLP11,MMT11a,BJMM12a}.
This can be done for the parameters suggested in \cite{LJ12a}. 

\section{Implementation of the attack for the parameters suggested in \cite{LJ12a}}

We have carried out the attack on the parameters suggested in \cite{LJ12a}.
They are provided in Table \ref{tab:parameters}.
\begin{table}[!h]
\caption{Parameters for the second scheme suggested in \cite{LJ12a}.}
\begin{center}
\begin{tabular}{|c|c|c|c|c|c|c|c|c|}
\hline 
$n$ & $n_{B}$ & $k$ &$k_{B}$ &$b$ & $c$ &$L$ & $m$ & $t$ (number of errors)  \\
\hline 1800&  1020 &  1160 & 660 & 20 & 30 & 25 & 12 & 45 \\ \hline 
\end{tabular}
\end{center}
\label{tab:parameters}
\end{table}%

\begin{figure}[!h]
\begin{center}
\includegraphics[scale=0.42]{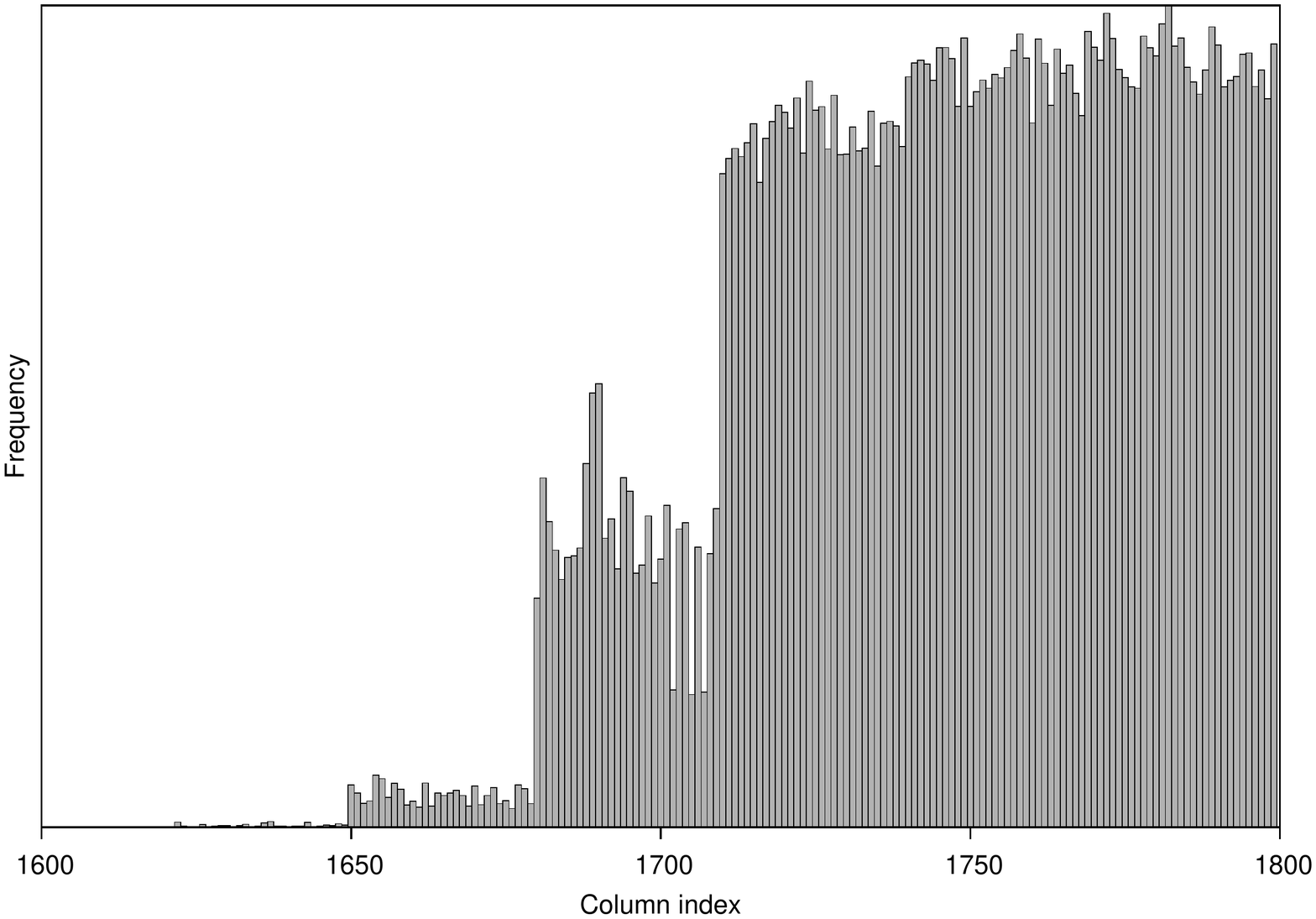}
\end{center}
\caption{The frequencies of the code positions involved in codewords of weight $\leq 22$ output by Dumer's algorithm.} \label{fig:Dumer22}
\end{figure}

Setting the weight parameter $w$ accurately when calling the function 
{\tt LowWeight} is the key for finding the $60$ last positions. If 
$w$ is chosen to be too large, for instance when $w=22$,  running  Dumer's low weight 
codeword search algorithm  \cite{Dumer91} gave the  result given in Figure \ref{fig:Dumer22}
concerning the frequencies of the code positions involved in the codewords of weight less than $22$ output by the algorithm and stored in table $tab$ during the execution of the algorithm.

\begin{figure}[!h]
\begin{center}
\includegraphics[scale=0.42]{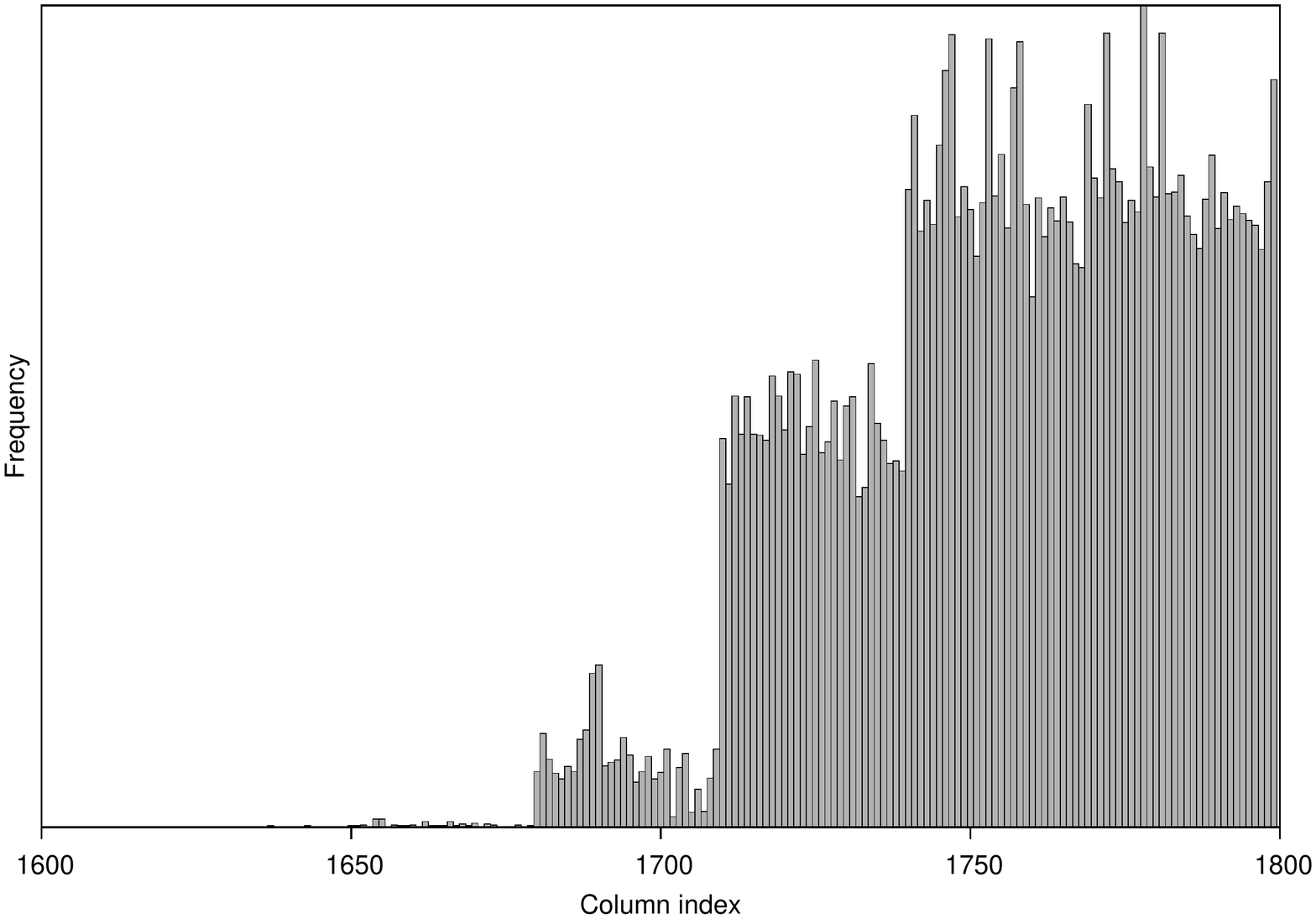}
\end{center}
\caption{The frequencies of the code positions involved in codewords of weight $\leq 18$ output by Dumer's algorithm.} \label{fig:Dumer18}
\end{figure}

 We see in Figure \ref{fig:Dumer22} that 
 this discriminates the $90$ last code positions and not as we want  the $60$ last code positions.
 However choosing $w$ to be equal to $18$ enables to discriminate the $60$ last positions as shown 
 in Figure \ref{fig:Dumer18}.
 
%% RAJOUTER UN BLABLA SUR l'ATTAQUE, temps effectif pour casser le Goppa
%% Estimation de la complexite
%% Mettre une reference bibliographique vers ton implémentation : tu l'as rendu publique non ?

Data used in Figure \ref{fig:Dumer18} come from $3900$ codewords generated in one hour and a half on an Intel Xeon W3550 (3 GHz) CPU by a monothread implementation in C of Dumer's algorithm.
The message recovery part of the attack involving the Goppa code consists in decoding  $25.5$  errors on average in a linear code of dimension $660$ and 
length $1020$. The time  complexity is about $2^{42}$. This second part of the attack could be achieved using the previous program on the same computer in about   $6.5$ hours on average.

% -*- mode: latex ; mode: auto-fill; mode: flyspell ; -*-
\section{Analysis of the security of the scheme} \label{sec:analysis}

\subsection{An improved attack}
\label{ss:improvement}

The purpose of this section is to provide a very crude analysis of the security of the scheme. 
We will not analyze our attack detailed in Section 
\ref{sec:idea}, since even if it was enough to break the second scheme suggested in \cite{LJ12a} it is not the most efficient one.
We will give a sketch of a better attack and a rough analysis for it. Basically, the real threat on this scheme
comes from the fact that there exists a subcode $\Cc$ of $\Cpub$ of very small support (of size $2c$ here), namely the code generated by the last 
$b$ rows of $\Gms$ permuted by the secret permutation matrix $\Pm$. For instance, there are about $2^{b-1}$ codewords of weight less
than or equal to $c$ which should be found by a low weight codeword searching algorithm and which should reveal
the support of $\Cc$. This is basically the idea underlying our attack. However there are other subcodes of rather small support
which yield low weight codewords, namely the codes $\Cc_{s}$ generated by the $s\times b$ last rows of $\Gm$ for $s$
ranging between $2$ and $L$.  The support of $\Cc_{s}$ is of size $(s+1)c$. Notice that its rate gets closer and closer to the rate
$\frac{2}{3}$ (which is more or less the rate of the final code) as $s$ increases. This is a phenomenon which helps low weight 
codeword  algorithms as will be explained later on.

An improvement of our attack would consist in using a low weight codeword algorithm in order to find one
of the codewords of $\Cc_{s}$ and to use this codeword $c$ to bootstrap from here to find the whole support of $\Cc_{s}$.
This is very much in the spirit of the attack against the KKS scheme which is explained in Algorithm 2 which can be found
in Subsection 4.4 of \cite{OT11}. With this approach, by using the codeword which has already been found, it is much easier to 
find new ones belonging to the same subcode with small support by imposing that the information set used for finding 
low weight codewords is chosen outside the positions belonging to the support of $c$. The complexity of the whole attack is dominated 
in this case by the complexity of finding just one codeword in $\Cc$ when there is a good way to identify the candidates 
in $\Cc$ (which can be done by checking the weight of $c$).  
Notice that it is very likely that $\Cc$ is actually the sub code of $\Cc$ of dimension 
$b$ which has the smallest support. Recall here that this is precisely the notion captured by the generalized Hamming weights of a code \cite{WY93},
$w_{i}$ being defined as the smallest support of a subcode of dimension $i$. In other words $w_{1}$ is nothing but the
minimum distance of the code and in our case it is likely that $w_{b}=2c$ (and more generally 
$w_{sb}=(s+1)c$ for $s =1..L$).
In other words, the problem which should be difficult 
to solve is the following one
\begin{problem}
Find one of  the subcodes of dimension $s \times b$ whose support size is the $s \times b$-th generalized Hamming weight
of $\Cpub$.
\end{problem}

We will focus now on  the following approach to solve this problem.
Consider a low weight codeword algorithm which aims at finding low weight codewords in a code of dimension $k$
by picking up a random set of positions $\Ic$ of size slightly larger than $k$, say $k+l$ and which looks for all (or at least a non-negligible
fraction) of codewords which have weight  equal to some small  quantity $p$ over these positions. These quantities
are very good candidates for having low weight over the whole support. This is 
precisely the approach which is followed in the best low weight codeword search algorithms such as 
\cite{Stern88,Dumer91,MMT11a,BJMM12a}. We run such an algorithm for several different sets $\Ic$ and will be interested in the
complexity of outputting at least one codeword which belongs to $\Cc$. This is basically the approach which has been very successful
to break the KKS scheme \cite{OT11} and which is the natural candidate to break the \cite{LJ12a} scheme.

%The complexity of such algorithms is of the form $O(\frac{C_{1}}{P})$, where $C_{1}$ is the complexity of checking one set $\Ic$ and
%$P$ is the probability that a codeword which has less than the target weight  is of weight $p$ over the set $\Ic$. Here the target weight can be set to
%$n'$ the length of $\Cc$.
%Before analyzing such a scheme, let us first explain why analyzing such an algorithm needs some special treatment. For instance, the crude
%approach which consists in writing that $P$ is approximated by $\sum_{w=1}^{n'} a_{w} P(n,k,w)$, where $a_{w}$ is the number of codewords of weight $w$ in $\Cc$ and $P(n,k,w)$ is the complexity of finding a codeword of weight $w$ in 
%a code of dimension $k$ and length $n$ (which are respectively the dimension and the length of $\Cpub$) when there is one
%such codeword by one of the aforementioned algorithm, say the best one which is in \cite{BJMM12a}, is wrong here. The error consists
%here in approximating the union of events $\prob\left(\cup_{c \in \Cc, c \neq 0} E_{c}\right)$, where $E_{c}$ is the event 
%``$c$ is output by the algorithm'' by the sum 
%$\sum_{c \in \Cc, c \neq 0} \prob(E_{c})$ which is poor in this case, since the probabilities of the joint events
%$\prob(E_{c}\cap E_{c'})$ is  far away from the product $\prob(E_{c})\prob(E_{c'})$.
To analyze such an algorithm we will make some simplifying assumptions
\begin{itemize}
\item The cost of checking one of those $\Ic$ is of order 
$O\left( L + \frac{L^2}{2^l}\right)$ where $L=\sqrt{\binom{k+l}{p}}$. We neglect here the cost coming from writing the parity-check matrix
in systematic form and this does not really cover the recent improvements in \cite{MMT11a,BJMM12a}. We have made here
such an approximation for sake of simplicity.
We refer to  \cite{FinSen09} for an explanation of this cost.
\item We assume that the result of the puncturing of $\Cc$ by all positions which do no belong to its support  behaves like a random code of dimension $k'$ and length $n'$.
\end{itemize}

Our main result to analyze such an algorithm consists in the following proposition.
\begin{proposition}
\label{pr:complexity}
Let 
\begin{itemize}
\item$f(x)$ be the function defined by $f(x) \eqdef \max \left(x(1-x/2),1-\frac{1}{x} \right)$ ;
\item $p(s) \eqdef \frac{\binom{n'}{s}\binom{n-n'}{k+l-s}}{\binom{n}{k+l}}$;
\item $\lambda(s) \eqdef  \binom{s}{p}2^{k'-s}$.
\item $C(k,l,p) \eqdef L + \frac{L^2}{2^l}$ where $L \eqdef \sqrt{\binom{k+l}{p}}$;
\item $P \eqdef \sum_{s=1}^{n'} p(s) f(\lambda)$.
\end{itemize}
Then the complexity that the low weight codeword search algorithm outputs an element in 
$\Cc$ is of order 
$$
O\left(\frac{ C(k,l,p)}{P} \right).
$$
\end{proposition}

\subsection{Proof of Proposition \ref{pr:complexity}}

Our first ingredient is a lower bound on the probability that a given set $X$ intersects a random linear code
$\Cr$ of dimension $k$ and length $n$ picked up uniformly at random. This lemma gives a sharp lower bound even when 
$X$ is very large and when there is a big gap between the quantities 
$\prob(X \cap \Cr \neq \emptyset)=\prob( \cup_{x \in X} \{x \in \Cr\} )$ and $\sum_{x\in X} \prob(x \in \Cr)$.

\begin{lemma}
\label{lem:lower_bound}
Let $X$ be some subset of $\Ft^n$ of size $m$ and let $f$ be the function 
defined by 
$f(x) \eqdef \max \left(x(1-x/2),1-\frac{1}{x} \right)$.
We denote by $x$ the quantity $\frac{m}{2^{n-k}}$, then 
$$
\prob(X \cap \Cr \neq \emptyset) \geq f(x).
$$
\end{lemma}

This lemma can be found in \cite{OT11} and it is proved there.

Let us finish now the proof of Proposition \ref{pr:complexity}. Denote by $\Jc$ the support of $\Cc$:
$$
\Jc \eqdef \suppt{\Cc}.
$$
Let us first calculate the expected number of 
sets $\Ic$ we have to consider before considering an element of $\Cc$. Such an event happens precisely when 
there is a nonzero word in $\Cc$ whose restriction to $\Ic \cap \Jc$ is of weight  equal to $p$.
Let $\Cc_{\Ic \cap \Jc}$ be the restriction of the codewords of $\Cc$ to the positions which belong to 
$\Ic \cap \Jc$, that is 
$$
\Cc_{\Ic \cap \Jc} \eqdef \{(c_{i})_{i \in \Ic \cap \Jc}: (c_{i})_{1 \leq i \leq n} \in \Cc\}.
$$
Let $X$ be the set of non-zero binary words of support $\Ic \cap \Jc$ which have weight equal to $p$.
Denote by $W$ the size of $\Ic \cap \Jc$. The probability that $W$ is equal to $s$ is precisely
$$
\prob(W=s) = \frac{\binom{n'}{s}\binom{n-n'}{k+l-s}}{\binom{n}{k+l}} = p(s).
$$
Then the probability $P$ that a certain choice of $\Ic$ gives among the codewords considered by the algorithm 
a codeword of $\Cc$ can be expressed as 
\begin{eqnarray}
P &= &\sum_{s=1}^{n'} \prob(W=s) \prob(X \cap \Cc' \neq \emptyset) \\
& \geq & \sum_{s=1}^{n'} p(s) f(\lambda)
\end{eqnarray}
by using Lemma \ref{lem:lower_bound} with $\Cc'$ and the aforementioned $X$.
Therefore the average number of iterations which have to be performed before finding an element in 
$\Cc$ is equal to $\frac{1}{P}$ and this yields immediately Proposition \ref{pr:complexity}.

\subsection{Repairing the parameters and a pitfall}
\label{ss:repair}

A possible way to repair the scheme consists in increasing the size of the random part (which corresponds to the last 
$c$ columns in $\Gms$ here). Instead of choosing this part to be of size $c$ as suggested in \cite{LJ12a}, its size can be 
increased in order to thwart the algorithm of Subsection \ref{ss:improvement}. Let $r$ be the number of random columns we add
at the end of the convolutional part, so that the final length of the code is now $n_{B}+Lc+r$ instead of $n_{B}+(L+1)c$ as before. 
If we choose 
$r$ to be equal to $140$, then the aforementioned attack needs about $2^{80}$ operations before outputting an 
element of $\Cc$ which is the (permuted) subcode corresponding to the last $b$ rows of $\Gms$.
As before, let us denote by $\Cc_{s}$ the permuted (by $\Pm$) subcode of $\Cpub$ generated by the
last $s \times b$ rows of $\Gms$ permuted by $\Pm$. We can use the previous analysis to estimate the complexity of obtaining 
an element of $\Cc_s$ by the previous algorithm. We have gathered the results in Table \ref{tab:complexity}.

\begin{table}[!h]
\caption{Complexity of obtaining at least one element of $\Cc_{s}$ by the algorithm of
Subsection \ref{ss:improvement}}
\begin{center}
\begin{tabular}{|c|c|c|c|c|c|c|c|c|}
\hline
$s$ & $1$ & $5$ &$10$ & $15$ & $20$ &$21$ &$22$&  $25$  \\
\hline complexity (bits) & $80.4$ & $72.1$ & $65.1$ & $61.0$ 
&$59.4$ & $59.3$ & $59.4$ & $59.8$ \\ \hline 
\end{tabular}
\end{center}
\label{tab:complexity}
\end{table}%

We see from this table that in this case the most important threat does not come from finding 
low weight codewords arising from codewords in $\Cc_{1}$, but codewords of moderate weight
arising from codewords in $\Cc_{20}$ for instance. Codewords in this code have average weight
$\frac{r+20c}{2}=370$. This implies that a simple policy for detecting such candidates which
consists in keeping all the candidates in the algorithm of Subsection \ref{ss:improvement} which have weight less than this quantity is very likely to filter
out the vast majority of bad candidates and keep with a good chance the elements of $\Cc_{20}$.
Such candidates can then be used as explained in Subsection \ref{ss:improvement} to check whether or not they belong 
to a subcode of large dimension and small support.

There is a simple way for explaining what is going on here. Notice that the rate of $\Cc$ is equal to $\frac{b}{c+r}$, which is much 
smaller than the rate of the overall scheme which is close to $\frac{b}{c}$ in this case by the choice of the parameters
of the Goppa code. However as $s$ increases, the rate of $\Cc_{s}$ gets closer and closer to $\frac{b}{c}$, since 
its rate is given by $\frac{sb}{sc+r}=\frac{b}{c+r/s}$. Assume for one moment that the rate of $\Cc_{s}$ is equal to 
$\frac{b}{c}$. Then putting $\Gmp$ in systematic form (which basically means that we run the aforementioned algorithm with $p=1$
and $l=0$) is already likely to reveal most of the support of $\Cc_{s}$ by looking at the support of the rows which have weight around
$\frac{sc+r}{2}$ (notice that this phenomenon was already observed in \cite{Ove07}).
This can be explained like this. We choose $\Ic$ to be of size $k$,  the dimension of $\Cpub$, and to be an information set for $\Cpub$.
Then, because the rate of $\Cc_{s}$ is equal to the rate of $\Cpub$, we expect that the size 
of $\Ic \cap \Jc$ (where $\Jc$ is the support of $\Cc_{s}$) has a rather good chance  to be 
of size smaller than or equal to the dimension of $\Cc_{s}$. This in turn implies that it is possible
to get codewords from $\Cc_{s}$ by any choice over the information set $\Ic$ of weight $1$ which is non zero 
over $\Ic \cap \Jc$ (and therefore of weight $1$ there). More generally, even if $\Ic \cap \Jc$ is slightly bigger
than the dimension of $\Cc_{s}$ we expect to be able to get codewords in $\Cc_{s}$ a soon as 
$p$ is greater than the Gilbert-Varshamov distance of the restriction $\Cc'_{s}$ of $\Cc_{s}$ to $\Ic \cap \Jc$, 
because there is in this case a good chance that this punctured code has codewords of weight $p$.
This Gilbert-Varshamov distance will be very small in this case, because the rate of $\Cc_{s}$ is very close to $1$
(it is expected to be equal to $\frac{\dim (\Cc_{s})}{|\Ic \cap \Jc|}$).

Nevertheless, it is clear that it should be possible to set up the parameters (in particular increasing $r$ should do the job) so that existing low weight 
codeword algorithms should be unable to find these subcodes $\Cc_{s}$ with complexity less than
some fixed threshold. However, all these codes $\Cc_{s}$ have to be taken into account and the attacks
on the dual have also to be reconsidered carefully (\cite{LJ12a} considered only attacks on the dual
aiming at finding the codewords of lowest weight, but obviously the same technique used for finding
some of the $\Cc_{s}$ will also work  for the dual). Moreover, even if by construction
the restriction of $\Cc=\Cc_{1}$ to its support should behave as a random code, this is not true anymore
for $\Cc_{s}$ with $s$ greater than one, due to the convolutional structure. The analysis sketched in Subsection \ref{ss:improvement}
should be adapted a little bit for this case and should take into account the improvements over 
low weight searching algorithms \cite{MMT11a,BJMM12a}.  Finally, setting up the parameters also requires a careful study of the error probability that sequential decoding fails. This whole thread of work is beyond the scope of the present paper.

\bibliographystyle{alpha}
\bibliography{codecrypto}

\end{document}